\documentclass[11pt,fleqn,twoside]{article}
\usepackage{amsmath,graphicx,bm,a4wide}

\usepackage{amsmath,graphicx,bm,cite}

\newcommand{\addressstyle}{\em\normalsize}
\newcommand{\figureandcaption}[3][0pt]{\begin{figure}[t]
\parbox[b]{.48\linewidth}{\includegraphics[width=\linewidth]{#2}}
\hfill\parbox[b]{.48\linewidth}{\caption{#3\vspace{#1}\ }}
\end{figure}}

\begin{document}

\title{Stationary State Fluctuation Theorems for Driven Langevin Systems}

\author{
E.\ G.\ D.\ Cohen \textsuperscript{a}
and 
Ramses van Zon\textsuperscript{a,b} 
\\\textsuperscript{a}\addressstyle
The Rockefeller University, 1280 York Avenue, New York, NY 10021-6399, USA
\\\textsuperscript{b}\addressstyle 
Chemical Physics Theory Group, Department of Chemistry, University of Toronto,
\\\addressstyle 80 St.\ George St., Toronto, ON M5S 1A1, Canada
}

\date{January 26, 2007}

\maketitle

\begin{abstract}
Recent results on the stationary state Fluctuation Theorems for work
and heat fluctuations of Langevin systems are presented. The relevance
of finite time corrections in understanding experimental and
simulation results is explained in the context of an exactly solvable
model, namely a Brownian particle in a harmonic potential, which is
dragged through the surrounding fluid. In this model, work
fluctuations obey the conventional form of the fluctuation theorem
while heat fluctuations satisfy an extended form. The connection with
other work in recent literature is pointed out, and further
generalizations are suggested.
\end{abstract}

\section{Introduction}

In the last twelve years, the field of non-equilibrium statistical
mechanics has been given a boost by the discovery of a number of new,
general results regarding the fluctuations of thermodynamic
quantities such as heat, work and entropy.  This paper is not intended
as a survey article, rather, we will restrict ourselves to
non-equilibrium stationary state Fluctuation Theorems (FTs).  A
stationary state FT was first discovered in \cite{Evansetal93}.  This
FT is a quantitative relation between the probability to observe a
positive entropy production $\sigma$ versus a negative one $-\sigma$
over a time interval $\tau$.  Given the probability distribution
$\pi_\tau$ of a fluctuating quantity $p$, scaled such that its average
is equal to one, the Conventional Fluctuation Theorem (CFT) (also
called the Gallavotti-Cohen Fluctuation Theorem) takes the
form\cite{GallavottiCohen95}
\begin{equation}
\lim_{\tau\to\infty} \frac{1}{\tau} \log
\frac{\pi_\tau(p)}{\pi_\tau(-p)}
=\sigma_+p \text{ for } p<p^*,
\label{CFT}
\end{equation}
where $\sigma_+$ is the average (dimensionless) entropy production
rate over positive times, $p=\sigma/\sigma_+$, and $p^*$ is an upper
bound on the values of $p$ for which the CFT holds.

Whether the CFT holds depends on the kind of system and on what the
quantity $p$ represents. Choices for $p$ that have been used in the
literature include the phase space contraction rate
\cite{GallavottiCohen95}, a ``dissipation
function''\cite{EvansSearles02}, an ``action
functional''\cite{LebowitzSpohn99}, and, more concretely, work, heat
and entropy production.  The value of $p^*$ can also depend on the
quantity under consideration. The stationary state Fluctuation Theorem
holds quite generally (i.e. for a large class of
deterministic\cite{Evansetal93, GallavottiCohen95,EvansSearles02} and
stochastic systems\cite{Kurchan98, LebowitzSpohn99}), and has been
verified in laboratory experiments\cite{Cilibertoetal04,
GarnierCiliberto05}.

The behavior for values $p>p^*$ has recently been elucidated for
specific systems, and led to an Extended Fluctuation Theorem
(EFT) for heat\cite{VanZonCohen03a, VanZonCohen03b04,
VanZonetal03e04}, which we will discuss in detail below.

In this paper, we will give an overview of some interesting results of
the stationary state Fluctuation Theorem for an exactly solvable model
of a Brownian particle in a harmonic potential which is dragged
through a surrounding fluid.  While this model is based on a
Langevin equation, the conclusions from this model also pertain to
certain purely deterministic systems, as we will point out.

We will first introduce the model in Sec.~\ref{nebp} and discuss its
exact solution in Sec.~\ref{solution}. We will then turn our attention
to thermodynamic quantities of the dissipated heat and the supplied
work, and their fluctuations, both for finite and infinite time in
Sec.~\ref{flucts}. In Sec.~\ref{FTs} we will show that this results in
two kinds of stationary state fluctuation relations: the Conventional
Fluctuation Theorem (CFT), for work fluctuations, and an Extended
Fluctuation Theorem (EFT), for heat fluctuations. The paper ends with
a discussion in Sec.~\ref{conclusion} of the results and
generalizations that have appeared in the literature.

\section{The model}
\label{nebp}

Consider a spherical Brownian particle of mass $m$ and diameter $D$
suspended in an equilibrium fluid with viscosity $\eta$ and
temperature $T$. The motion of this Brownian particle can be described
by the Langevin equation
\begin{equation}
 m  \ddot{\bm x} = -\alpha  \bm v + \bm{\xi},
\label{LE-equilibrium}
\end{equation}
where $\bm x$ is the position of the particle, $\bm v=d\bm x/dt$ its
velocity and $\alpha=3\pi\eta D$ the Stokes friction
coefficient. Furthermore, $\bm{\xi}=\bm{\xi}(t)$ is a Gaussian
fluctuating force whose average (over realizations)
$\langle\bm{\xi}(t)\rangle=0$ and whose time correlations are
delta-correlated, i.e.
\begin{equation}
  \langle \bm{\xi}(t_1)\bm{\xi}(t_2)\rangle
= 2k_BT \alpha\delta(t_1-t_2)\mathbf{1},
\label{noise-correlations}
\end{equation}
where $k_B$ is Boltzmann's constant and $\mathbf{1}$ is the identity
matrix.

Consider now the situation in which an external force acts on the
Brownian particle (but not on the fluid). Under the assumption that
the fluid is unperturbed such that $\alpha$ and the properties of
$\bm\xi$ are not changed, the Langevin equation gets modified to
\begin{equation}
 m \ddot{\bm x} = \bm{F}(\bm x,t)-\alpha\bm v + \bm{\xi},
\label{LE-non-equilibrium}
\end{equation}
where $\bm{F}(\bm x,t)$ is the external force, which may depend on
position and time.

Of particular interest here is the case in which the force derives
from a moving harmonic potential $U$, i.e.,
\begin{equation}
   \bm{F}(x,t)=-\frac{\partial U(\bm x,t)}{\partial\bm x} = -\kappa\bm(\bm x-\bm x^*(t)\bm),
\label{harmonic-force}
\end{equation}
where the potential $U(\bm x,t)=\kappa|\bm x-\bm x^*(t)|^2/2$ and the
center $\bm x^*(t)$ of the potential changes in time according to
\begin{equation}
  \bm x^*(t) = \bm v^* t.
\label{center-motion}
\end{equation}
This corresponds to the particle being trapped in a harmonic well, the
center $\bm x^*$ of which moves with a constant velocity $\bm v^*$, to
which we will restrict ourselves here. This closely resembles the
experimental system of a Brownian particle in a moving optical trap
studied by Wang \emph{et al.}\cite{Wangetal02}.

One also needs to specify the initial condition of the system.  In
time, the distribution of $\bm x$ and $\bm v$ approach a
non-equilibrium stationary state distribution $f_{\rm ness}$
asymptotically. It is not hard to show that
\begin{equation}
  f_{\rm ness}(\bm x,\bm v)=
  \frac{\beta\sqrt{\kappa m}}{2\pi}
\exp\left[ -\frac{\beta (m|\bm v-\bm v^*|^2+\kappa|\bm x-\bm v^*(t-\alpha/\kappa)|^2)}{2}
			 \right],
\label{fness}
\end{equation}
where $\beta=1/(k_BT)$.  Note that if $\bm v^*=0$ this reduces to the
equilibrium distribution.  To focus on the non-equilibrium stationary
state fluctuations, one can take as an initial condition that the
distribution at time $t=0$, denoted by $f_0$, is equal to $f_{\rm ness}$.

This system exhibits many features of general non-equilibrium systems,
namely the presence of a current, energy dissipation, and entropy and
heat production due to the continuous input of work from the outside.
Thus it can serve as a prototypical non-equilibrium system.  The
advantage of looking at this particular system is that it can be
solved exactly (as was first exploited for work fluctuations by
Mazonka and Jarzynski\cite{MazonkaJarzynski99}). In fact, this system
can also be solved exactly for a general the time dependence of
the center $\bm x^*(t)$ of the harmonic potential, as was shown in
\cite{VanZonCohen02b}.

\section{The exact solution}
\label{solution}

The exact solution of the Brownian particle system introduced above
proceeds through a time-dependent change of coordinates that brings
the equations into a form in which the force does not depend on time,
and whose solution is known.

As explained in \cite{VanZonCohen02b}, to eliminate the time
dependence of the force, such a coordinate transformation should take
the form
\begin{equation}
  \bm X(t) = \bm x(t)-\tilde{\bm x}(t),
\label{co-moving}
\end{equation}
where $\tilde{\bm x}$ is any solution of the equations of motion in
the absence of the fluctuating force. One such solution is
\begin{equation}
  \tilde{\bm x}(t) = \bm v^* (t-\alpha/\kappa).
\label{co-moving2}
\end{equation}
Note that the new coordinate system corresponds to a frame that is
co-moving with the fictitious noiseless trajectory $\tilde{\bm x}(t)$,
and that along this fictitious trajectory, the friction force $-\alpha
\bm v^*$ is exactly balanced by the harmonic force $-\kappa(\tilde{\bm
x}-\bm x^*)=\alpha\bm v^*$.

In the co-moving frame, the Langevin equation becomes [cf.\
Eqs.\,\eqref{LE-non-equilibrium}--\eqref{co-moving2}]
\begin{equation}
 m \ddot{\bm X} = -\kappa\bm X -\alpha \bm V + \bm\xi,
\label{LE-non-equilibrium-co-moving}
\end{equation}
where $\bm V=d\bm X/dt=\bm v-\bm v^*$. Equation
\eqref{LE-non-equilibrium-co-moving} is of the same form as
Eq.\,\eqref{LE-non-equilibrium} but with $\bm v^*=0$, so it describes
the system with a stationary harmonic potential. This system is
well-studied and is an example of a so-called Ornstein-Uhlenbeck
process\cite{UhlenbeckOrnstein}. Roughly speaking, an
Ornstein-Uhlenbeck process is a stationary Markov process that can be
described by a Langevin equation with linear forces and a Gaussian
random force term. Equation \eqref{LE-non-equilibrium-co-moving} is
Markovian because the time-derivatives of $\bm X$ only depend on
quantities at that same time. As a consequence of the Markovian nature
of the process, all its properties can be expressed in terms of the
Green's function $G_t(\bm X',\bm V'; \bm X,\bm V)$ which is the
probability, in the co-moving frame, for the particle to be at
position $\bm X'$ with velocity $\bm V'$ given that it was at position
$\bm X$ with velocity $\bm V$ a time $t$ earlier. For
Ornstein-Uhlenbeck processes, the Green's function can be solved
exactly and is Gaussian in both the initial coordinates ($\bm X$, $\bm
V$) and the final coordinates ($\bm X'$,$\bm V'$). It is this Gaussian
property that enables one to obtain many explicit results, as we will
see.

Although this system can be solved exactly, to avoid unnecessarily
complicated expressions, we will consider here only the overdamped
case, which means taking the limit $m\ll\alpha^2/k$. In
Eq.\,\eqref{LE-non-equilibrium-co-moving}, this limit can be taken by
setting $m$ equal to zero, so that it reduces to $ 0 = -\kappa\bm
X-\alpha \bm V + \bm\xi$, or
\begin{equation}
  \dot{\bm X} = -\frac{1}{\tau_r}\bm X + \alpha^{-1}\bm\xi,
\label{LE-non-equilibrium-co-moving-overdamped}
\end{equation}
where $\tau_r=\alpha/\kappa$ is the (average) relaxation time of this
Langevin equation. Like the general case, this equation also describes
an Ornstein-Uhlenbeck process, whose Green's function is given
by\cite{UhlenbeckOrnstein}
\begin{equation}
  G_t(\bm X'; \bm X) =
  \left[\frac{\beta\kappa}{2\pi(1-e^{-2t/\tau_r})}\right]^{1/2}
	\exp\left[ -\frac{\beta\kappa}{2}
				\frac{|\bm X'-e^{-t/\tau_r}\bm X|^2}
				{1-e^{-2t/\tau_r}}
			 \right].
\label{Greens}
\end{equation}

Given a set of times $0<t_1<t_2<\dots<t_m$, one can write for the
joint probability of the particle to be at position $\bm X_n$ at time
$t_n$ (with $t_0=0$):
\begin{equation}
  P(\bm X_0,\bm X_1,\ldots,\bm X_m) = f_0(\bm X_0)
 \prod_{i=1}^m G_{t_i-t_{i-1}}(\bm X_i; \bm X_{i-1}) .
\label{Ppath}
\end{equation}
Because each Green's function is Gaussian in both the initial and
final positions [cf.~Eq.\,\eqref{Greens}], and the initial
distribution in Eq. \eqref{fness} is also Gaussian in the initial
position $\bm X_0$, the distribution $P$ is also Gaussian.

\section{Thermodynamic fluctuations}
\label{flucts}

\subsection{Definitions}

While on a macroscopic level, thermodynamic properties such as work,
heat and entropy are well-defined, this is not the case on a
mesoscopic level. To understand that there is an ambiguity, consider
trying to determine the ``energy of the system'' for the Brownian
particle under consideration. Should this include the potential energy
of the external harmonic force, or should this only be the (kinetic)
energy of the Brownian particle itself? Or consider the ``work done on
the system''. Should this be the work done by the harmonic force on
the particle, or should this be the work needed to move the harmonic
potential?  The first choice is closely related to the mechanical
definition of work, but it leads to work being done even in
equilibrium (although its average would be zero), a problem that is
avoided by using the second choice.  There are even more ambiguities
when studying dissipation in non-equilibrium systems.  When should
energy be regarded as dissipated?  It would be ambiguous to say that
the dissipated energy is that part of the energy input that cannot be
used to perform work. For instance, the harmonic potential energy can
be used to store energy by displacing the Brownian particle from its
origin. Is this dissipated energy?  A clever experimentalist may be
able to extract useful work from this energy, but should the
definition of dissipated energy depend on the ability of an
experimentalist?

The ambiguity is thus due to what to call the system, and what forms
of energy to consider as inaccessible or dissipated.  To avoid this
ambiguity, we will now define what we will mean by ``work'' and
``heat'' in the rest of the paper. The work will be defined as the
work that is required to keep the harmonic potential moving. Because
at time $t$ the harmonic potential exerts a force $-\kappa\bm(\bm
x(t)-\bm x^*(t)\bm)$ on the particle, the particle will exert a
reaction force $\kappa\bm(\bm x(t)-\bm x^*(t)\bm)$ on the harmonic
potential (or on whatever device generated it). To keep the potential
moving therefore requires (at least) a compensating force
$-\kappa\bm(\bm x(t)-\bm x^*(t)\bm)$ to be applied to the
potential. Since the velocity of the potential is $\dot{\bm x}^*=\bm
v^*$ (cf. Eq~\eqref{center-motion}), the mechanical work done by this
compensating force during a time interval $\tau$ is
\begin{equation}
  W_\tau = -\kappa\bm v^*\cdot \int_0^\tau \!dt\, [\bm x(t)-\bm x^*(t)].
\label{Wdef}
\end{equation}
Note that this definition of ``work on the system'' uses the particle
plus the harmonic potential as the system.

On the other hand, we will define the heat as the energy which is
given off to the fluid, taking the point of view that since the energy
of the fluid is not explicitly present in the Langevin description, we
can suppose it to be unaccessible or dissipated energy. This leads to
the following expression for the heat:
\begin{equation}
  Q_\tau = W_\tau - \Delta U_\tau
\label{Qdef}
\end{equation}
where $\Delta U_\tau=U\bm(\bm x(\tau)\bm)-U\bm(\bm x(0)\bm)$ is the
change in harmonic potential from time $0$ to time $\tau$. In words,
this states that work can be converted both into potential energy of
the Brownian particle, or into heat dissipated into the fluid.  Note
that in the general (non-overdamped) case, the change in kinetic
energy of the particle should also be subtracted from $W_\tau$ to
obtain the heat.

\subsection{Fourier transforms}

Given the definitions of the work and the heat and the Green's
function of the process, we can study the probability distributions of
their fluctuations. Once the probability distributions are obtained,
it can be checked whether they obey a stationary state Fluctuation
Theorem. In addition (and in contrast to the majority of theoretical
works on the subject), \emph{the explicit solution available for this
system allows us to consider finite time corrections to these theorems
as well. These corrections give information about the time scales on
which these stationary state Fluctuation Theorems can be observed.}

An essential ingredient in the calculation will be that any linear
function of a Gaussian distributed quantity is again a Gaussian
distributed quantity, so that the distribution of such a quantity is
fully determined by its mean and its variance.  While the work
$W_\tau$ in Eq.\,\eqref{Wdef} is a linear function of $\bm x(t)$,
making it Gaussian distributed, the
heat $Q_\tau$ in Eq.\,\eqref{Qdef} is not a linear function of
$\bm x(t)$, but is in fact a quadratic function of $\bm x(t)$ and
consequently not Gaussian distributed. This quadratic nature prevents
a straightforward calculation of the heat probability distribution
function, but still allows the computation of its Fourier transform,
using the following trick.

Consider first the joint probability $P_\tau^*(W_\tau,\bm x(0),\bm
x(\tau))$ which gives the probability that the particle moves from
position $\bm x(0)$ to position $\bm x(\tau)$ and an amount $W_\tau$
of work is done in the process. Since these three quantities are all
linear functions of $\bm x(t)$, their joint distribution is Gaussian,
i.e. of the form
\begin{equation}
 P_\tau^*(\bm y) = \frac{e^{-\frac12(\bm y-\bm M)\cdot\mathsf
 V^{-1}\cdot (\bm y-\bm M)}}{\sqrt{2\pi\det \mathsf V}}
\end{equation}
where for brevity we denoted $\bm y=\bm(W_\tau, \bm x(0),
\bm x(\tau)\bm)$, $\bm M$ as the mean of $\bm y$ and $\mathsf V$ as
the variance matrix of $\bm y$.

Using $P^*_\tau$, the distributions of $W_\tau$ and $Q_\tau$ can be
expressed as
\begin{eqnarray}
  P_\tau^W(W) &=&
  \int\!d\bm y\:  P_\tau^*(\bm y)  \, \delta(W-W_\tau)
\\
  P_\tau^Q(Q) &=&
  \int\!d\bm y\:  P_\tau^*(\bm y)\,
  \delta\left(Q-W_\tau+\kappa|\bm x(\tau)-\bm x^*(\tau)|^2/2
  -\kappa|\bm x(0)-\bm x^*(0)|^2/2 \right)
,
\end{eqnarray}
where $\bm x^*(t)=\bm v^*t$ as given by Eq.~\eqref{center-motion}.
Taking the Fourier transform of these equations yields
\begin{eqnarray}
  \hat P_\tau^W(q) &=&
  \int\!d\bm y\:  P_\tau^*(\bm y)   e^{iq(W-W_\tau)}
\\
  \hat P_\tau^Q(q) &=&
\int\!d\bm y\:  P_\tau^*(\bm y)
  e^{iq\left(Q-W_\tau+\kappa|\bm x(\tau)-\bm v^*\tau|^2/2
  -\kappa|\bm x(0)|^2/2 \right)}
,
\end{eqnarray}
where $q$ is the Fourier variable.  The integrals on the right hand
sides are all just Gaussian integrals, which can be performed
explicitly once $\bm M$ and $\mathsf V$ are determined. The results
for the stationary state are
then\cite{VanZonCohen02b,VanZonCohen03b04}:
\begin{eqnarray}
  \hat P_\tau^W(q) &=& e^{\sigma_+\tau\tilde q \left[ i - \tilde
  q\left(1-\frac{1-e^{-\tau/\tau_r}}{\tau/\tau_r}\right) \right]}
\label{PW}
\\
  \hat P_\tau^Q(q) &=& 
  \frac{e^{\sigma_+\tau\tilde q\left[ i - \tilde q\right]\left[ 1 -
	\frac{2\tilde
	  q^2(1-e^{-\tau/\tau_r})}{\tau/\tau_r(1+(1-e^{-2\tau/\tau_r})\tilde q^2)}\right]}
  }{[1+(1-e^{-2\tau/\tau_r})\tilde q^2]^{3/2}}
\label{PQ}
.
\end{eqnarray}
where $\tilde q=k_BTq$ is the dimensionless Fourier variable and
$\sigma_+=\alpha\beta |\bm v^*|^2$ is the average dimensionless heat
production rate.

\subsection{Saddle point approximations and singularities} 

To obtain the actual probability distributions of work and heat
requires the inversion of the Fourier transforms $\hat P_\tau^W$ in
Eq.\,\eqref{PW} and $\hat P_\tau^Q$ in Eq.\,\eqref{PQ}. Since $\hat
P_\tau^W$ is Gaussian, its inverse is easily obtained:
\begin{equation}  
  P_\tau^W(W) = \frac{
                 e^{
                    -\frac{
                           (\beta W-\sigma_+\tau)^2
                     }{
                           4\sigma_+\tau\left(
                                            1-\frac{
                                                    1-e^{-\tau/\tau_r}
                                              }{
                                                    \tau/\tau_r
                                              }
                                      \right)
                     }
                  }
             }{
                 \sqrt{
                       2\pi \sigma_+\tau\left(
                                       1-\frac{
                                               1-e^{-\tau/\tau_r}
                                         }{
                                               \tau/\tau_r
                                         }
                                  \right)/\beta^2
                 }
             }
\end{equation}
In fact, it is not necessary to use the Fourier transform to obtain
this result for $P_\tau^W$, but it will be useful to have the Fourier
transforms of both the work and the heat distribution to understand
the origin of the peculiarities of the behavior of the heat
fluctuations.

The quantity of interest for the stationary state Fluctuation Theorems
to be derived in Sec. 5 is not the probability of the work
fluctuations but that of the \emph{scaled} work fluctuations $p=\beta
W/(\sigma_+\tau)$ (which compares the dimensionless $W_\tau$, $\beta
W_\tau$ with the average dimensionless entropy production over a time
$\tau$). Its probability distribution $\pi^W_\tau(p)$ is related to
that of $W$ by a Jacobian:
$\pi^W_\tau(p)=P_\tau^W(W)dW/dp=P_\tau^W\bm(\beta
W/(\sigma_+\tau)\bm)(\sigma_+\tau/\beta)$, so that
\begin{equation}  
\pi^W_\tau
  (p) = \frac{
                 e^{
                    -\frac{
                           \sigma_+\tau(p-1)^2
                     }{
                           4\left(
                                            1-\frac{
                                                    1-e^{-\tau/\tau_r}
                                              }{
                                                    \tau/\tau_r
                                              }
                                      \right)
                     }
                  }
             }{
                 \sqrt{
                       2\pi \left(
                                       1-\frac{
                                               1-e^{-\tau/\tau_r}
                                         }{
                                               \tau/\tau_r
                                         }
                                  \right)/(\sigma_+\tau)
                 }
             } 
\label{PiW}
\end{equation}

The Fourier inversion of $\hat P_\tau^Q$ is more involved because no
closed form is known for the Fourier inverse of Eq.\,\eqref{PQ}. We
are therefore left with the following formal expression for
$\pi^Q_\tau(p)$:
\begin{equation}
  \pi^Q_\tau(p) = \frac{\sigma_+\tau}{2\pi}\int_{-\infty}^{\infty}
 d\tilde q \:e^{-i\tilde q\sigma_+\tau p}
  \frac{e^{\sigma_+\tau\tilde q\left[ i - \tilde q\right]\left[ 1 -
	\frac{2\tilde
	  q^2(1-e^{-\tau/\tau_r})}{\tau(1+(1-e^{-2\tau/\tau_r})\tilde q^2)}\right]}
  }{[1+(1-e^{-2\tau/\tau_r})\tilde q^2]^{3/2}}
\label{PiQ}
.
\end{equation}
We are especially interested in the large $\tau$ behaviour of this
function. The presence then of a large parameter in the exponent in
the integrand allows us to obtain an approximate result using the
saddle point approximation. To exhibit the large parameter $\tau$ in
Eq.\,\eqref{PiQ} explicitly, we write
\begin{equation} 
  \pi^Q_\tau(p) = \frac{\sigma_+\tau}{2\pi}\int_{-\infty}^{\infty}
d\tilde q \:e^{\tau h_\tau^Q(\tilde q)},
\end{equation}
where
\begin{equation} 
  h_\tau^Q(\tilde q) =
\sigma_+\left\{-i\tilde q p+\tilde q\left[ i - \tilde q\right]\left[ 1 -
	\frac{2\tilde
	  q^2(1-e^{-\tau/\tau_r})}{(\tau/\tau_r)(1+(1-e^{-2\tau/\tau_r})\tilde q^2)}
\right]
\right\}
-\frac3{2\tau}
\log\left[
 1+(1-e^{-2\tau/\tau_r})\tilde q^2
\right].
\end{equation}

In the saddle point method, the main contribution to the integral is
found by considering the maximum $\tilde q^*$ of the function
$h_\tau^Q(\tilde q) $. A complicating factor is that $h_\tau^Q$ is
complex valued. If one takes $\tilde q^*$ to be the point on the real
line where $h_\tau^Q$ has a maximum real part, then the integrand can
be highly oscillatory there, causing the contribution of this
otherwise dominant point to be washed out.  A maximum that does not
suffer from this oscillatory behavior can be found, but usually not on
the real axis; indeed the solution of
\begin{equation}  
  \frac{dh_\tau^Q(\tilde q^*)}{d\tilde q^*} = 0
\label{saddlepoint}
\end{equation}
lies here in the complex plane.

An interesting property of complex functions is that they have no true
maximum. According to the Cauchy-Riemann identities for complex
functions, the point $\tilde q^*$ defined by \eqref{saddlepoint} has
the property that there is a path through it called the path of
steepest descent, on which $h_\tau^Q$ has a constant imaginary part
and its real part has a maximum, while along another (orthogonal) path
the function has a minimum (also with constant imaginary part); thus
$\tilde q^*$ is a saddle point of $h_\tau^Q$.  The saddle point
character of $\tilde q^*$ does not interfere with it being the
dominant contribution to the integral, as long as we can deform the
original line of integration (the real axis) into a contour which
overlaps with the path of steepest descent through $\tilde
q^*$. Furthermore, it can be shown that along this line the function
does not oscillate, so that the point $\tilde q^*$ truly gives the
dominant contribution. Note that during the deformation of the
integration contour, the endpoints ($\pm\infty$ on the real axis) must
be fixed and no singularities of the function must be crossed. The
value of integral along this deformed contour is then exactly equal to
that of the original integral. To get the saddle point approximation,
one expands the function $h_\tau^Q(q)$ around $\tilde q^*$ up to
second order in $\delta\tilde q=\tilde q-\tilde q^*$, and integrates
along the deformed contour. As $\tau$ increases, the function becomes
more peaked around the saddle point, and the saddle point expression
becomes an increasingly good approximation to the integral.

To illustrate this method, let us first apply it to the work
distribution, which is simpler than the heat distribution and for
which the result can be compared with the exact result in
Eq.\,\eqref{PiW}. From Eq.\,\eqref{PW} one then obtains
\begin{equation}  
  \pi^W_\tau(p) = \frac{\sigma_+\tau}{2\pi}\int_{-\infty}^{\infty}
d\tilde q \:e^{\tau h_\tau^W(\tilde q)},
\end{equation}
where
\begin{equation} 
  h_\tau^W(\tilde q) =
\sigma_+\left\{-i\tilde q p+\tilde q\left[ i - \tilde q
\left(
1-\frac{1-e^{-\tau/\tau_r}}{\tau/\tau_r}
\right)
 \right]
\right\}.
\label{hW}
\end{equation}
Since this is a quadratic function, the saddle point $\tilde q^*$ can
be obtained analytically from $dh_\tau^W/d\tilde q^*=0$, with the
result
\begin{equation}  
   \tilde q^* = i \frac{1-p}{2\left(
1-\frac{1-e^{-\tau/\tau_r}}{\tau/\tau_r}
\right)
}.
\label{qstarW}
\end{equation}
Note that the saddle point lies on the imaginary axis. Noting also
that $h_\tau^W(\tilde q)$ has no singularities as a function of
$\tilde q$, one can deform the integration contour (i.e. the real
axis) to go through $\tilde q^*$. Around $\tilde q^*$ the function
$h_\tau^W$ has the following behavior:\footnote{For more general forms
of $h$, there would be $O(\delta\tilde q^3)$ correction terms.}
\begin{equation}  
  h_\tau^W(\tilde q^*+\delta\tilde q) = -\sigma_+\left\{
\frac{(1-p)^2}{4
  \left(
1-\frac{1-e^{-\tau/\tau_r}}{\tau/\tau_r}
\right)
}
  +\delta \tilde q^2
\left(
1-\frac{1-e^{-\tau/\tau_r}}{\tau/\tau_r}
\right)
\right\}.
\end{equation}
Note that for real $\delta\tilde q$, this expression is real (i.e.,
non-oscillatory) and has a maximum at $\delta\tilde q=0$: the path of
steepest descent therefore crosses $\tilde q^*$ parallel to the real
axis.  Thus we can indeed deform the original line of integration (the
real axis) by shifting the integration contour up or down (depending
on $p$) until it crosses $\tilde q^*$; this changes the end points
$\pm\infty$, but the integrand vanishes there, so that the integral
over the shifted line is still the same as for the original
integration line. This is sketched in left panel of figure~1. The
final integral now becomes
\begin{equation}  
  \pi^W_\tau(p) = \frac{\sigma_+\tau}{2\pi}e^{ -
\frac{\sigma_+\tau(1-p)^2}{4
  \left(
1-\frac{1-e^{-\tau/\tau_r}}{\tau/\tau_r}
\right)
}}
\int_{-\infty}^{\infty}
d\delta\tilde q \:e^{-\sigma_+\tau 
\left(
1-\frac{1-e^{-\tau/\tau_r}}{\tau/\tau_r}
\right)\delta \tilde q^2}.
\end{equation}
After performing the Gaussian integral, one recovers Eq.\,\eqref{PiW}.

We now turn to the distribution of heat fluctuations. To find the
saddle point in this case, the derivative of $h_\tau^Q(\tilde q)$ has
to be equal to zero. This derivative is rather complicated, in
contrast to the simple linear function that it was for the work
functions. To make the expressions in the following a bit more
manageable, let us assume that $\tau$ is so large that
$e^{-\tau/\tau_r}$ may be neglected in Eq.\,\eqref{saddlepoint} (by
comparison with numerical methods, we found that this typically leads
to accurate results for $\tau> 3\tau_r$\cite{VanZonCohen03b04}). In
that case, $h_\tau^Q$ becomes
\begin{equation}  
  h_\tau^Q(\tilde q)  =
\sigma_+\left\{
-i\tilde q p+\tilde q( i - \tilde q)
+\frac{2\tilde
	  q^3}{(\tau/\tau_r)(i+\tilde q)}
\right\}
-\frac3{2\tau} \log( 1+\tilde q^2 ).
\label{approxhQ}
\end{equation}
while
\begin{equation}  
 \frac{d}{d\tilde q} h_\tau^Q(\tilde q)  =
\sigma_+\left\{-i p + i -2\tilde q +
	\frac{6\tilde q^2}{(\tau/\tau_r)(i+\tilde q)}
+
	\frac{2\tilde q^3}{(\tau/\tau_r)(i+\tilde q)^2}
\right\}
- \frac{3\tilde q}{\tau(1+\tilde q^2) }.
\label{dhQdq}
\end{equation}
Note that the naive limits $\tau\to\infty$ of $h_\tau^Q$ in
Eq.\,\eqref{approxhQ} and $h_\tau^W$ in Eq.\,\eqref{hW}
coincide. However, the terms in Eq.\,\eqref{approxhQ} that are then
neglected have singularities, which have to be taken into account,
because even though the factor $1/\tau$ may tend to zero as
$\tau\to\infty$, it multiplies an expression that can, because of the
singularities, go to infinity, making the limit $\tau\to\infty$
ill-defined.

\figureandcaption[5mm]{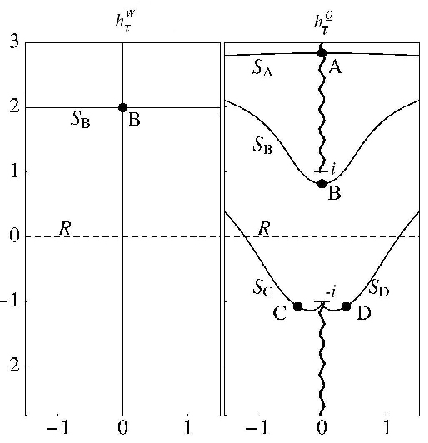} {Saddle points and paths of
steepest descent of the functions $h_\tau^W$ (left) and $h_\tau^Q$
(right) for $\sigma_+=1$, $\tau=4\tau_r$ and $p=-2$ (see text). Solid
circles represent saddle points and solid curves $S_\mathrm A,\ldots,
S_\mathrm D$ drawn through these are the paths of steepest
descent. For $h_\tau^W$, shown on the left, only one saddle point B
exists and the (dashed) real axis $R$ can be deformed to the indicated
path of steepest descent $S_{\rm B}$ going horizontally through $B$,
without crossing a singularity. For $h_\tau^Q$ shown on the right,
wiggly lines represent branch cuts, two singularities appear at $\pm
i$, and there are four saddle points A, B, C and D. The path of
steepest descent $S_{\rm B}$ through B is the (only) way to deform the
real axis to a path of steepest descent without crossing a
singularity.\label{fig1}}

The saddle point equation $dh_\tau^Q/d\tilde q=0$ leads, using
Eq.\,\eqref{dhQdq}, to a fourth order polynomial equation. Such an
equation has four solutions; hence there are four saddle points, in
contrast to the Fourier transform of the work distribution, which had
only one.  Which saddle point(s) to use depends on how we can deform
the real axis integration contour to include saddle points without
crossing any singularities. As Eq.\,\eqref{approxhQ} shows, there are
singularities at $\tilde q=\pm i$. The point $\tilde q=i$ is a branch
point associated with the logarithm, while $\tilde q=-i$ is in
addition a simple pole. The structure of the saddle points is shown in
the right panel of figure \ref{fig1}. As a function of $p$, there is
only one saddle point that we can reach by deforming the real axis
without crossing singularities. This saddle point lies on the
imaginary axis and is in a sense trapped between the singularities at
$\pm i$; this is quite different from the saddle point for the work
fluctuations [i.e. $i(1-p)/2$ for large $\tau$,
cf. Eq.\,\eqref{qstarW}] which is unbounded as a function of $p$.

The full calculation of the Fourier inverse of $\hat P_\tau^Q$ would
be too detailed for this short paper and can be found in
\cite{VanZonCohen03b04}. In an asymptotic expansion for large $\tau$,
the result is
\begin{equation}  
  \pi^Q_\tau(p) =
\begin{cases}   
\sqrt{\frac{\sigma_+^3\tau^3|p+1|}{36\pi}}
e^{-\sigma_+[\tau_r-\tau p]+3/2}
&
\text{ if } p<-1.
\\
\sqrt{\frac{16\sigma_+\tau}{\pi}}
\frac{e^{-\sigma_+(1-p)^2[\tau+2(1-p)/(3-p)\tau_r]/4}}
{[(3-p)(1+p)]^{3/2}}
&
\text{ if } -1<p<3.
\\
\sqrt{\frac{\sigma_+\tau^2/\tau_r}{32\pi}}
e^{-\sigma_+\left[(p-2)\tau-\sqrt{8(p-3)\tau\tau_r}
+\frac{20-6p}{p-3}\tau_r
\right]}
&
\text{ if } p>3.
\end{cases}
  \label{finitetauQ}
\end{equation}
Note that this function has a Gaussian center for $-1<p<3$ and
exponential tails for $p<-1$ and $p>3$, i.e., for large $\tau$, one
has $\pi_\tau\sim \exp[\sigma_+\tau p]$,
$\pi_\tau\sim\exp[-\sigma_+\tau(p-1)^2/4]$ and $\pi_\tau\sim
\exp[-\sigma_+\tau (p-2)]$ for $p<-1$, $-1<p<3$ and $p>3$,
respectively.  Qualitatively, this result can be understood as
follows: if $p$ is such that the saddle point $\tilde q^*$ is
relatively far away from the singularities, the function and its
relevant saddle point resemble those of the work fluctuations, i.e.,
$\tilde q^*$ scales linearly with $p$ and the Fourier inverse is
consequently Gaussian in $p$. Since for the work fluctuations $\tilde
q^*=i(1-p)/2$, the values of $p$ for which $\tilde q^*$ is not close
to the singularities $\pm i$ lie within the interval $p\in[-1,3]$. For
values $p>3$, the work saddle point lies below $-i$, but the heat
saddle point has to stay within $[-i,i]$. As a result it is basically
stuck just above $-i$, i.e., $\tilde q^*\approx -i$, yielding,
cf. Eq.\,\eqref{PiQ}, an exponential dependence of $\pi_\tau^Q(p)$ on
$p$. Likewise, for $p<-1$, the heat saddle point is stuck just below
$i$ and again an exponential tail of the distribution is present.

Eq.\,\eqref{finitetauQ} also includes finite $\tau$ effects, obtained
using the saddle point approximation. Note that the approach to the
large $\tau$ behavior has a different character for different values
of $p$. This corresponds to the different character of the
singularities at $\pm i$ (i.e., pole vs. branch point).

\section{Stationary State Fluctuation Theorems}
\label{FTs}

Given the distribution functions, one can now investigate whether the
stationary state Fluctuation Theorem holds.  The quantity considered
in the Fluctuation Theorems is the \emph{fluctuation function}
$f_\tau(p) = \frac{1}{\sigma_+\tau}
\log[\pi_\tau(p)/\pi_\tau(-p)]$. This function measures the asymmetry
between positive and negative fluctuations in the quantity $p$. When
$p$ represents work, $\pi_\tau(p)=\pi^W_\tau(p)$ and Eq.\,\eqref{PiW}
gives:
\begin{equation}  
   f^W_\tau(p) = \frac{p}{1-\frac{1-e^{-\tau/\tau_r}}{\tau/\tau_r}} .
\label{fW}
\end{equation}
As $\tau\to\infty$, this tends to $p$ from above (cf.\ figure~2),
hence the Conventional Fluctuation Theorem (CFT)is satisfied by the
work fluctuations. For these fluctuations, the characteristic time is
$\tau_r$.

On the other hand, considering the heat fluctuations and using
Eq.\,\eqref{finitetauQ}, one finds an Extended Fluctuation Theorem
(EFT):
\begin{equation}  
   f^Q_\tau(p) = \begin{cases}
                 p + \frac{8p}{(9-p^2)\tau/\tau_r}-\frac{3\ln\frac{3-p^2+2p}{3-p^2-2p}}{2\sigma_+\tau}
                 & \text{ if } 0<p<1
                 \\
                 p - \frac{(1-p)^2}{4} 
		 + \frac{(p+1)(-2p^2+8p-10)}{4(3-p)\tau/\tau_r}
		 - \frac1{\sigma_+\tau}\left\{
                 \ln(\sigma_+\tau)
                 -\frac12\ln \frac{(3-p)^3(1+p)^3(p-1)}{576}
                 -\frac32
		 \right\}
                 & \text{ if } 1<p<3
                 \\
                 2 + \sqrt{\frac{8(p-3)}{\tau/\tau_r}} -
                 \frac{\ln(\tau\sigma^2\tau_r)}{2\sigma_+\tau} 
                 & \text{ if } p>3
                 \end{cases}
\label{fQ}
\end{equation}
Note that the behavior for negative $p$ can simply be found from the
antisymmetry of the function $f^Q_\tau(p)$,
i.e. $f^Q_\tau(-p)=-f^Q_\tau(p)$.

To illustrate the difference between the work and heat fluctuation
functions $f_\tau^W$ and $f_\tau^Q$, we sketched these functions as a
function of $p$ in figure~\ref{fig2}.  One sees that in the case
$\tau\to\infty$, for $p<1$, the fluctuation functions tend to a
straight line with slope one. This behavior persists beyond $p>1$ for
$f^W_\tau$, while $f^Q_\tau$ bends down away from the straight line
and settles to a plateau value of $2$ for $p>3$.

Thus, even in the limit $\tau\to\infty$, Eq.~\eqref{fQ} for heat
fluctuations differs from the CFT for work fluctuations when $p>1$,
since the fluctuation function has for heat a plateau at a value of
$2$ rather than that it increases without bound as for
work. Alternatively, one could say that $f_\tau^Q$ does satisfy
Eq.\,\eqref{CFT}, but with $p^*=1$, whereas $f_\tau^W$ has
$p^*=\infty$.  We note that for the heat fluctuations there is in
addition to $\tau_r$ a relaxation time $\sigma_+^{-1}$, which can
exceed $\tau_r$.

\figureandcaption[5mm]{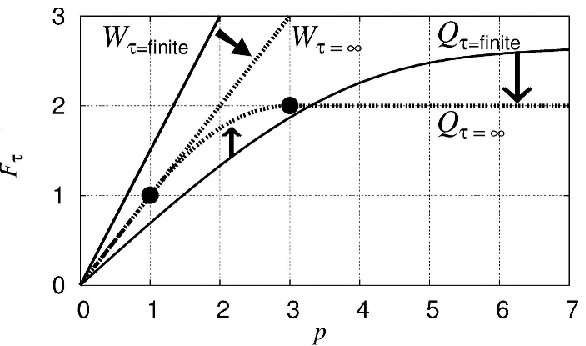} {Conventional (CFT) and Extended
Fluctuation Theorem (EFT). $W_{\tau=\rm finite}$ and $W_{\tau=\infty}$
represent the CFT and $Q_{\tau=\rm finite}$ and $Q_{\tau=\infty}$ the
EFT for finite and infinite times, respectively.  The arrows indicate
how the infinite time limit is approached, i.e., for the work
fluctuations from above, and for the heat fluctuations from below for
small $p$ and from above for large~$p$.\label{fig2}}

The plateau value 2 in the EFT for $p>3$ (cf. Eq.~\eqref{fQ}) is a
consequence of the exponential tails of the distribution
$P_\tau^Q$. These tails are related to the presence of singularities
at $\pm i$, which limit the position of the saddle point $\tilde q^*$,
such that for $p>3$ one has $\tilde q^*\approx -i$, while for $p<-1$,
$\tilde q^*\approx +i$. As the saddle point gives the dominant
contribution to the Fourier inverse, thus leads to the exponential
tails
\begin{equation}
\begin{array}{rcll}
\pi_\tau^Q(p)&\sim& e^{-\sigma_+\tau (p-2)} &\text{ for } p>3
\\
\pi_\tau^Q(p) &\sim& e^{\sigma_+\tau p} &\text{ for } p<-1.
\end{array}
\end{equation}
The exponential tails directly give
$f^Q_\tau(P)=(\sigma_+\tau)^{-1}\ln\pi_\tau^Q(p)/\pi_\tau^Q(-p)\sim
2$.  Note that conversely, exponential tails give rise to
singularities in the Fourier transform since $\int_{-\infty}^\infty
dp\, e^{iqp} e^{-|p|} = i/(q+i)-i/(q-i)$.

\section{Discussion}
\label{conclusion}

We have shown that for a dragged Brownian particle described by a
Langevin equation, the work fluctuations satisfy the Conventional
Fluctuation Theorem, and the heat fluctuations satisfy an Extended
Fluctuation Theorem. These results may seem to only be peculiarities
of this model. However, these results largely carry over to other more
general systems too.

1. The model of a Brownian particle in a harmonic potential dealt with
here can be mapped onto a system of an electric circuit with a current
source\cite{VanZonetal03e04}. Furthermore, the model can be extended
to the Rouse model for a polymer, which consists of a harmonically
bound system of Brownian particles. It has been shown that there too
the work distribution is Gaussian and satisfies the
CFT\cite{SpeckSeifert05}, while the heat fluctuations appear to have
exponential tails\cite{Dhar05}.

2. Another extension is to consider the non-overdamped case. This has
been done in the context of work fluctuations by Douarche \textit{et
al.}\cite{Douarchetal06}. While the details of the work fluctuations
for finite times can be quite different, the general conclusion,
namely, that the work fluctuations satisfy a CFT, still holds.  It is
hard to imagine that the heat would not satisfy an EFT, since this
model is just a more detailed description of the same Brownian system.

3. Extensions to non-harmonic potentials have been explored as
well. Blickle {\it et al.} have studied the work fluctuations in
\cite{Blickleetal06}, while Baiesi {\it et al.}\cite{Baiesietal06}
have argued for an EFT for heat under some general assumptions, one of
which is the existence of exponential tails for the distribution of
$Q_\tau$.

4. The limitation $p<p^*$ to the CFT has been found analytically by
other authors as well\cite{Bonettoetal06a,ReyBelletThomas02}. The
limit is in all these cases also due to singularities in the complex
plane of the Fourier transform of the distribution, which correspond
to exponential tails of the distribution itself.

5. Turning to deterministic systems, Evans\cite{Evans05} argued that
there should be a relation between the FT in such systems and the
EFT. Indeed, Gilbert has shown numerically that the EFT holds in a
Nose-Hoover thermostated Lorentz Gas\cite{Gilbert06} and
\cite{Bonettoetal06a} explains in a general way why the EFT should
hold using a large deviation formalism, but still assuming that there
are exponential tails and that the work fluctuations have faster than
exponential tails. They also provide a strategy on how to apply the
chaotic hypothesis to singular systems such as Lennard-Jones systems,
by using a Poincar\'e section that circumvents the singularities in
the potential.

6. A derivation of a general form of the EFT has been given for
stochastic systems by Baiesi \textit{et al.} \cite{Baiesietal06} and
for deterministic systems by Bonetto \textit{et al.}
\cite{Bonettoetal06a}, both based (among others) on the presence
of exponential tails. They found that universal features in the EFT of
the $\lim_{\tau\to\infty}f^Q_{\tau}(p)$ are the slope one for $p<1$
and the presence of a plateau for large $p$. On the other hand, the
shape and extent of the intermediate-$p$ region is not universal, nor
is the height of the plateau, unless the work distribution is
symmetric around its average, in which case the plateau value is
always 2.

7. Evidence for the EFT has been seen in experiments as well, notably
by Garnier and Ciliberto in an electric
circuit\cite{GarnierCiliberto05}.

8. It has been suggested that the Gallavotti-Cohen Fluctuation Theorem
(Eq.~(1)), meaning the stationary state Fluctuation Theorem for
entropy production, or heat, would not hold near equilibrium, but the
FT for the ``dissipation function'' of Evans and Searles (which often
coincides with work) would\cite{EvansSearlesRondoni05}.  However, it
seems that in the appropriate limit $\tau\to\infty$, with a
restriction $p<p^*$, the Gallavotti-Cohen Fluctuation Theorem does
hold. Nonetheless, there is a problem in observing the EFT for heat
near equilibrium that the CFT for work does not have. As we can see in
the exactly solved model of the Brownian particle, the FT for work
decays to the correct FT form on a time scale of $O(\tau_r)$
[cf. Eq.\,\eqref{fW}]. One the other hand, the heat FT for $p<1$
decays to the correct FT on a time scale of $O(\sigma_+^{-1})$
[cf. Eq.\,\eqref{fQ}]. The latter time scale diverges as $\sigma_+$
gets smaller, i.e., as one gets closer to equilibrium, whereas the
time scale for the work FT (i.e., $\tau_r$) does not. This explains
why it is very hard to see the FT for heat in
simulations\cite{FTdifficult2C}.

\section*{Acknowledgments}

This work was supported by grants from the Department of Energy's
Office of Basic Engineering under grant DE-FG-02-88-ER13847 and the
National Science Foundations' Mathematics-Physics Division under grant
PHY-0501315.

\end{document}